\documentclass[aps, reprint, prd, superscriptaddress]{revtex4-2}

\usepackage[utf8]{inputenc}
\usepackage[T1]{fontenc}
\usepackage{booktabs, array, mathptmx, float, tabularx, booktabs, lipsum, amsmath,multirow}

\usepackage{siunitx, xcolor}
\usepackage{hyperref}
\usepackage{orcidlink}
\hypersetup{linkcolor=red,citecolor=green,filecolor=cyan,urlcolor=magenta}

\newcommand{\obs}{_\mathrm{o}}
\newcommand{\coa}{_0}
\newcommand{\realcoa}{_\mathrm{c}}
\newcommand{\ee}{\mathrm{e}}
\newcommand{\fourier}[1]{\mathcal{F}\left\{#1\right\}}

\usepackage{graphicx}
\usepackage{physics}
\usepackage{upgreek}

\DeclareSIUnit{\parsec}{pc}

\begin{document}

	\title{A Novel Method to Construct Frequency-Domain Gravitational Waveform for Accelerating Sources}
	\author{Xinmiao Zhao\orcidlink{0000-0002-5651-4356}}
	\affiliation{Department of Astronomy, School of Physics, Peking University, Beijing 100871, People’s Republic of China}
	\author{Han Yan\orcidlink{0000-0003-1320-5243}}
	\affiliation{Department of Astronomy, School of Physics, Peking University, Beijing 100871, People’s Republic of China}
	\author{Xian Chen\orcidlink{0000-0003-3950-9317}}
	\email{xian.chen@pku.edu.cn}
	\affiliation{Department of Astronomy, School of Physics, Peking University, Beijing 100871, People’s Republic of China}
	\affiliation{Kavli Institute for Astronomy and Astrophysics, Peking University, Beijing 100871, People’s Republic of China}
	
\begin{abstract}
Accurately modeling the inspiral--merger--ringdown (IMR) signal of coalescing
compact objects is essential for the test of general relativity.  However, it is known
that astrophysical environments can distort gravitational-wave (GW) signal and, if ignored, may
bias parameter estimation or even our understanding of gravity.  Previous
studies suggest that various astrophysical environmental effects can be modeled
in a unified way by introducing an effective acceleration. However, such models are 
based on
stationary phase approximation (SPA) and post-Newtonian (PN) formalism, which
are inconsistent with the fast orbital evolution and strong gravity in the
final merger--ringdown phase.  To overcome this limit, we introduce frequency-domain
spectral differentiation (FSD), which maps the time shift of the signal caused
by acceleration into a differentiation in the frequency domain.  The mapping
does not rely on SPA or PN formalism, therefore can be used to construct the accelerated waveform
across the entire IMR phases.  We compare the FSD waveforms with the
conventional SPA+PN ones, and find that the former more faithfully match the
simulated signals of accelerating sources, especially in the merger--ringdown
phase and when  higher-order FSD corrections are included.  A
Fisher information matrix analysis suggests that FSD waveforms can achieve
higher precision than SPA+PN waveforms in measuring effective acceleration.
Therefore, the FSD method offers a more self-consistent treatment of various  astrophysical
environmental effects in the final merger--ringdown phase of binary GW sources.
\end{abstract}
	
	\maketitle

	\section{Introduction}\label{sec:intro}

Gravitational-wave (GW) astronomy has entered an era of precision physics and
astrophysics, marked by a rapidly growing number of high signal-to-noise-ratio
(SNR) events from LIGO–Virgo–KAGRA (LVK) observations
\citep{ligoscientificcollaborationandvirgocollaborationGWTC1GravitationalWaveTransient2019,abbottGWTC2CompactBinary2021,theligoscientificcollaborationGWTC3CompactBinary2021,collaborationGWTC40ConstraintsCosmic2025},
as well as the prospect of having next-generation higher-sensitivity detectors in
the next decade
\citep{punturoEinsteinTelescopeThirdgeneration2010,reitzeCosmicExplorerUS2019}.  One of
the ultimate goals of precise GW observation is to test the robustness of
general relativity (GR) and the nature of compact objects in the regime of highly-dynamical strong gravity
\citep{willConfrontationGeneralRelativity2009,bertiTestingGeneralRelativity2015,cardosoTestingNatureDark2019,yunesGravitationalwaveTestsGeneral2025}.
Although the current data from LVK have not shown evidence of alternative
gravitational theories
\citep{abbottTestsGeneralRelativity2019,ligoscientificcollaborationandvirgocollaborationTestsGeneralRelativity2021,theligoscientificcollaborationTestsGeneralRelativity2021,collaborationGWTC40TestsGeneral2026a,collaborationGWTC40TestsGeneral2026,collaborationGWTC40TestsGeneral2026b},
they do reveal subdominant waveform features, such as higher modes in the
inspiral-merger phase
\citep{ligoscientificcollaborationandvirgocollaborationGW190412ObservationBinaryblackhole2020,abbottGW190814GravitationalWaves2020a,millsMeasuringGravitationalwaveHigherorder2021}
and quasi-normal modes in the ringdown part of the waveform
\citep{isiTestingNoHairTheorem2019,isiTestingBlackHoleArea2021,capanoMultimodeQuasinormalSpectrum2023,theligoscientificcollaborationBlackHoleSpectroscopy2026,yangContributionNonlinearQuasinormal2025,wangNonlinearVoiceGW2501142026},
which are useful in the test of non-GR theories
\citep{kasthaTestingMultipoleStructure2018,pangPotentialObservationsFalse2018a,breschiIMRConsistencyTests2019,gieslerBlackHoleRingdown2019,mitmanNonlinearitiesBlackHole2023,cheungNonlinearEffectsBlack2023,mehtaTestsGeneralRelativity2023,mahapatraMultiparameterMultipolarTest2024}.

To test GR, the current data-analysis pipelines often implicitly assume that
the GW source, such as a binary of compact objects, evolves
in vacuum and in isolation
\citep{veitchParameterEstimationCompact2015,romero-shawBayesianInferenceCompact2020}. 
In reality, compact objects form and coalesce in complex
astrophysical environments, where the surrounding matter and
massive bodies can interact with the GW source and perturb the waveform
(e.g. \cite{barausseCanEnvironmentalEffects2014,bonvinEffectMatterStructure2017,zwickPrioritiesGravitationalWaveforms2023}).
If these environmental imprints are neglected or inaccurately modeled, particularly
during the critical stages such as  merger and ringdown, severe biases 
may be incurred in determining the correct model. For example,
recent studies have demonstrated that unmodeled astrophysical
systematics can mimic signatures
of modified gravity or non-linear gravitational dynamics \citep{sainiSystematicBiasParametrized2022,narayanEffectIgnoringEccentricity2023,guptaPossibleCausesFalse2025,gargSystematicsTestsGeneral2024,kejriwalHierarchicalModelingGravitationalwave2026}.
Therefore, it is urgent to prepare waveform models that incorporate astrophysical environmental
effects, to serve as the basis for future diagnosis of genuine non-GR signatures. 

Acceleration is a common environmental effect on GW signal
\cite{bonvinEffectMatterStructure2017}. To induce large enough deviation in
waveform, the body which is causing the acceleration could be a stellar-mass
compact object
\cite{antoniniBinaryBlackHole2017,meironDETECTINGTRIPLESYSTEMS2017,xuanDegeneracyMassPeculiar2021} or a
supermassive black hole (SMBH)
\cite{antoniniSECULAREVOLUTIONCOMPACT2012,bonvinEffectMatterStructure2017,inayoshiProbingStellarBinary2017,wongBinaryRadialVelocity2019}.
In terms of data analysis, acceleration results in a time-dependent Doppler
shift of the GW frequency and an aberration of the wave vector
\cite{torres-orjuelaPhaseShiftGravitational2020}, both of which can be modeled by a
phase shift in the frequency domain
\cite{chamberlainFrequencydomainWaveformApproximants2019,vijaykumarWaltzingBinariesProbing2023a,lazarowGravitationalWaveformModel2024a,tiwariProfilingStellarEnvironments2025,tiwariPipelineSearchSignatures2025,postiglioneEvolutionLISAObservables2025}.
Application of such waveform template to real LVK data has revealed interesting
candidates \cite{yangIndicationCompactObject2025}, but the significance is relatively low and
the effect is difficult to distinguish from those induced by higher multipolar modes, orbital 
precession, or eccentricity 
 \cite{tiwariPipelineSearchSignatures2025}. Besides acceleration-induced phase shift, 
a nearby third body can also
secularly perturb the orbital elements of a binary GW source, and in this way
affects the waveform \cite{antoniniSECULAREVOLUTIONCOMPACT2012,naozEccentricKozaiLidovEffect2016a,fangImpactSpinningSupermassive2019,kuntzEffectiveTwobodyApproach2023}. However, this effect is less
important during the merger and ringdown phase because of the short duration.

Gas surrounding a GW source is another environmental factor which can
modify the GW signal. For example, it has been shown that if a stellar-mass binary black hole (BBH)
forms in a dense stellar envelope or the accretion disk of an active galactic nucleus (AGN),
various dynamical processes, such as hydrodynamical friction \cite{barausseCanEnvironmentalEffects2014,fedrowGravitationalWavesBinary2017,tagawaFormationEvolutionCompactobject2020,liHydrodynamicalSimulationsBlack2023,whiteheadGasAssistedBinary2024}, gas torque \cite{dorazioOrbitalEvolutionEqualmass2021,liOrbitalEvolutionBinary2021,liHydrodynamicalEvolutionBlackhole2022,liHydrodynamicalEvolutionBlack2023,liHydrodynamicalEvolutionBlackhole2024,zwickNovelCategoryEnvironmental2024,zwickDissectingEnvironmentalEffects2025},
and accretion onto the BBH \cite{caputoGravitationalwaveDetectionParameter2020,cardosoEccentricityEvolutionCompact2021},
can modify the orbital evolution of the binary, and in turn modulate the phase of the waveform.
If not accounted for, these gas effects could
bias parameter estimation \cite{chenFakeMassiveBlack2020} or
 be misinterpreted as non-GR signatures \cite{royCompactBinaryCoalescences2025,canevasantoroFirstConstraintsCompact2024}.
To distinguish the gas effects, previous studies often rely on detailed simulations of the
gas dynamics \cite{liOrbitalEvolutionBinary2021,liHydrodynamicalEvolutionBlackhole2022,rowanBlackHoleBinary2023,whiteheadGasAssistedBinary2024,dittmannEvolutionInclinedBinary2024,dittmannMultiplePathsMerger2025}, which is computationally expensive and difficult to generalize.
It is worth mentioning that dark matter can also exert friction on an embedded binary and induce similar modulations on
GW signal \cite{barausseCanEnvironmentalEffects2014,edaGravitationalWavesProbe2015,yueDarkMatterEfficient2019,cardosoConstraintsAstrophysicalEnvironment2020,kavanaghDetectingDarkMatter2020,coleMeasuringDarkMatter2023,boudonGravitationalWavesBinary2024},
although the distribution of dark matter around GW source is relatively unclear.

These previous works suggest that a large class of environmental effects
modulate GW phase.  Therefore, recent studies focus on deriving different
phase-shift formulae to be used in different cases (e.g.,
\citep{riveraMeasurableParameterCombinations2024,zwickNovelCategoryEnvironmental2024,takatsyConstructionUseDephasing2025}).
However, due to the equivalence to frequency change, the net effect of phase
shift on waveform is similar to a Doppler shift, albeit some time dependence in
real astrophysical systems.  This analogy has led to the suggestion that
environmental effect can be modeled in a unified way by the waveform of an
accelerating source
\citep{chenFakeMassiveBlack2020,xuanDegeneracyMassPeculiar2021,chenDistortionGravitationalWaveSignals2021}. 

Meanwhile, searching for accelerating BBHs in LVK catalog is becoming an important
direction in GW astronomy
\cite{vijaykumarWaltzingBinariesProbing2023a,yangIndicationCompactObject2025,collaborationGWTC40TestsGeneral2026}.
To speed up the search, the phase shift induced by acceleration is usually
analytically calculated and added to the frequency-domain waveform.  Such
analytical phase shift is commonly calculated based on a stationary phase
approximation (SPA), which assumes that the wave frequency evolves slowly
compared to the orbital time-scale
\cite{chamberlainFrequencydomainWaveformApproximants2019,vijaykumarWaltzingBinariesProbing2023a,lazarowGravitationalWaveformModel2024a,tiwariPipelineSearchSignatures2025}.
Slow frequency evolution is valid during the early inspiral phase. Since at
this stage the orbital velocity is much smaller than the speed of light $c$,
the orbital dynamics has been modeled using post-Newtonian (PN) formalism with
sufficient accuracy \cite{buonannoComparisonPostNewtonianTemplates2009}.

The above framework of modeling accelerating GW source, however, may not apply
to the late inspiral, merger and ringdown waveforms.  On one hand, the high orbital velocity
in the late inspiral renders the PN formalism insufficient.  On the other, the highly
nonlinear dynamics during the merger phase and the multiple ringdown modes also
make the SPA invalid. Given the essential role of the
late-inspiral--merger--ringdown phase in testing GR
\citep{bertiTestingGeneralRelativity2015,yunesGravitationalwaveTestsGeneral2025,fooSystematicBiasDue2024},
new methods, independent of SPA or PN approximation, must be devised to
disentangle environment-induced acceleration-like effects from genuine non-GR
ones. 

To overcome the above difficulty, we consider an accelerating source, and map
the stretching of GW signal in the time domain to the phase shift in the
Fourier domain. We show that mathematically the time shift caused by
acceleration translates into a differentiation in the frequency domain.  This
operation is computationally fast, and does not depend on SPA or PN
approximation, therefore can be applied to the state-of-the-art
inspiral--merger--ringdown (IMR) models, such as IMRPhenomXPHM and SEOBNRv5PHM
\citep{prattenComputationallyEfficientModels2021,ramos-buadesNextGenerationAccurate2023}.

The paper is organized as follows. In Sec. \ref{sec:method}, we introduce the
traditional time-domain stretching method and develop a new Frequency-domain
Spectral Differentiation (FSD) framework. In Sec. \ref{sec:result}, we
compare the accelerated waveforms calculated using our method
against time-domain stretching and SPA-based approaches. In Sec.
\ref{sec:discussion}, we summarize the differences and prospect of our method.
Throughout the paper, we use geometric units where $G=c=1$.

\section{Method} \label{sec:method}

\subsection{General Consideration}

We aim to address the limitation of current waveform models for accelerating
sources, which rely on the SPA and thus fail
to accurately capture the merger and ringdown phases of the signal. To overcome
this issue, we first establish the relationship between
vacuum waveforms and accelerated waveforms in the time domain. We then present
two frequency-domain methods for constructing accelerated waveforms, in
Sections~\ref{sec:tds} and \ref{sec:fsd}, respectively.

As noted in the introduction, additional phase shifts in GW waveform can arise
from various processes, but they can be consistently modeled as a variation in redshift, $1+z(t)$. 
Change of redshfit corresponds effectively to a change in
line-of-sight velocity by $\Delta v\simeq cz(t)$ (assuming $z\ll1$). 
This common relationship, therefore,  allows the phase shifts to be unified
within the framework of an accelerating source.
Expanding the effective redshift around a reference time $t=t\coa$ at which
the redshift is $z\coa$,
we get
\begin{equation}
	1+z_\mathrm{eff}(t) \simeq 1 + z\coa + \left(\dv{z_\mathrm{eff}}{t}\right)\coa (t-t\coa).\label{eq:zeff}
\end{equation}
For BBHs in the sensitive band of 
ground-based detectors, it is customary to choose  $t\coa$ as
the time of coalescence $t\realcoa$ \citep{bonvinEffectMatterStructure2017,lazarowGravitationalWaveformModel2024a},
although alternative time points can be
selected depending on the specific requirements or convenience of the problem
at hand.
To relate the redshift change to an acceleration, we 	
define the effective line-of-sight acceleration as
\begin{equation} 
a =
-\frac{1}{2}\frac{\left(\dv*{z_\mathrm{eff}}{t}\right)\coa}{1+z\coa},
\label{eq:a_def} 
\end{equation}
where we have used the geometric units and 
the factor $-1/2$ is introduced for the simplification of the later equations.

Notice that the above first-order expansion is accurate only when $|a(t-t\coa)| \ll 1$.
This condition is normally valid for the BBHs detected by ground-based interferometers.
For example, the signal is typically much shorter than $1~{\rm s}$ for the current
LIGO/Virgo/KAGRA detectors \citep{ligoscientificcollaborationandvirgocollaborationObservationGravitationalWaves2016,ligoscientificcollaborationandvirgocollaborationGWTC1GravitationalWaveTransient2019}.
For the future more sensitive detectors,
such as the Cosmic Explorer \citep{reitzeCosmicExplorerUS2019} or Einstein
Telescope \citep{punturoEinsteinTelescopeThirdgeneration2010},
the duration of the signal could elongate to $10^2~{\rm s}$ \citep{maggioreScienceCaseEinstein2020,borhanianListeningUniverseNext2024}.
If the BBH is close to a SMBH
(e.g. \cite{bonvinEffectMatterStructure2017,meironDETECTINGTRIPLESYSTEMS2017,inayoshiProbingStellarBinary2017,chenExtrememassratioInspiralsProduced2018})
the resulting acceleration is approximately
\begin{equation}
		a_\mathrm{N} \sim \frac{M}{r^2}= \SI{2.0e-5}{\per\s} \left(\frac{M}{10^6 M_\odot}\right)^{-1}\left(\frac{r}{100\, r_\mathrm{g}}\right)^{-2}, \label{eq:Newtonian_a}
\end{equation}
where $M$ is the mass of the SMBH, $r$ is the distance between the SMBH and the
center-of-mass of the  BBH, and $r_\mathrm{g}=M$ is the gravitational radius of
the SMBH. Therefore, over the course of the signal's duration, the product $|a(t-t\coa)|$
remains much smaller than unity, unless we consider more massive SMBHs or closer distances.

\subsection{Time Domain Stretching} \label{sec:tds}

The redshift formulated in Equation~(\ref{eq:zeff}) modulates the time when the signal is observed.
Consequently, the elapsed proper time $\mathrm{d} \tau$ in the rest frame of the source is connected to the
elapsed time $dt$ in the observer's frame by the relation
\begin{equation}
		\dv{\tau}{t} = \frac{1}{1+z_\mathrm{eff}} \simeq \left(1 + z\coa\right)^{-1}\left[1+2a \left(t-t\coa\right)\right].
\end{equation}
This relationship allows us to stretch or squeeze the waveform in the time domain and accurately derive the
observed GW signal. In the following,
for convenience we set $1+z\coa=1$, and the above relationship simplifies to
\begin{equation}
	\tau - \tau\coa = \left(t - t\coa\right) +a\left(t-t\coa\right)^2,\label{eq:tau}
\end{equation}
where $\tau\coa$ is the reference time in the rest frame of the GW source.
Assuming that the amplitude $h\obs(t)$ of the observed GW signal is the same as the amplitude $h(\tau)$ in the rest frame 
of the source, which effectively means that we have neglected the effects such as gravitational lensing
or Doppler beaming, we can derive the observed waveform as
\begin{equation}
		h\obs(t) = h(\tau) = h\left(t - \Delta t\coa + a\left(t-t\coa\right)^2\right), \label{eq:ho}
\end{equation}
where $\Delta t\coa = t\coa - \tau\coa$.
	
The waveform templates routinely employed by the LVK
collaboration are constructed directly in the frequency domain, where the
frequency is defined in the source's rest frame. To incorporate the effects of
acceleration, our procedure begins by converting these standard
frequency-domain templates into the time domain via an inverse fast Fourier
transform (iFFT), yielding the signal $h(\tau)$. Subsequently, for each value
of the source's proper time $\tau$, we compute the corresponding observed time $t$
in the detector frame using Equation~(\ref{eq:tau}). This mapping from $\tau$ to $t$
results in a non-uniformly sampled signal in the observer's time. We
therefore interpolate this signal to obtain a uniformly sampled time series as
a function of $t$. Finally, a fast Fourier transform (FFT) is applied to this
interpolated data to produce the accelerated waveform in the frequency domain.

While this method of constructing accelerated waveforms, which we refer to as
``time-domain stretching'' (TDS), is in principle exact, it is computationally
expensive. To enhance efficiency, we therefore seek an approximation of
Equation~\eqref{eq:ho} that can be applied directly within the frequency domain. 
The details of the approximate method is presented in the following section.

\subsection{Frequency Domain Spectral Differentiation} \label{sec:fsd}
	
Before deriving the GW waveform in the frequency domain, we first
expand Equation~(\ref{eq:ho}) to the first order of $a(t-t\coa)^2$,
\begin{equation}
		\begin{split}
			h\obs(t) &= h\left(t - \Delta t\coa + a\left(t-t\coa\right)^2\right) \\
			&\simeq h(t - \Delta t\coa) + a\left(t-t\coa\right)^2 \pdv{h(t-\Delta t\coa)}{t}.
		\end{split}\label{eq:ho_approx}
\end{equation}
Here we have assumed that $|a(t-t\coa)^2\pdv*[2]{h(t-\Delta t\coa)}{t}|\ll
|\pdv*{h(t-\Delta t\coa)}{t}|$, which is equivalent to requiring that
$|a(t-t\coa)^2|\ll f_o^{-1}$, where $f_o$ is the frequency of the observed
waveform.  The validity of this assumption depends on two conditions: a
sufficiently small effective acceleration and a short time duration. These
conditions are typically met in the context of BBH detections with
next-generation ground-based interferometers.  For example, for a typical BBH
signal with a characteristic frequency of $\sim \SI{100}{\hertz}$ lasting
approximately $\SI{10}{\s}$, the assumption holds as long as the effective
acceleration satisfies $a< \SI{e-4}{\per\s}$.  This limit can be satisfied
according to the acceleration estimated in Equation~(\ref{eq:Newtonian_a}).
	
Next we perform a Fourier transformation on Equation~\eqref{eq:ho_approx}. 
We assume that the signal is
continuous with an infinite duration, so that 
Fourier transformation can be applied. 
After the transformation, the
observed signal in the frequency domain is 
\begin{equation} 
\begin{split}
\tilde{h}\obs (f_o) &= \fourier{h\obs(t)} \\ &\simeq \fourier{h(t - \Delta
t\coa)} + a\fourier{\left(t-t\coa\right)^2 \pdv{h(t-\Delta t\coa)}{t}} \\ &=
\tilde{h}(f)\ee^{-i\omega \Delta t\coa} + a\fourier{\left(t-t\coa\right)^2
\pdv{h(t - \Delta t\coa)}{t}}. 
\label{eq:ho_deriv} 
\end{split} 
\end{equation}
Here, $\mathcal{F}\{\cdots\}$ denotes the Fourier transformation,
\begin{equation}
	\mathcal{F}\{h(t)\} = \int_{-\infty}^\infty h(t)\ee^{-i\omega
t}\dd{t}, 
\end{equation} 
where $\omega = 2\uppi f$. It is important to notice that the frequency $f$ is defined in the
rest frame of the source, as it originates from the Fourier transform of the
waveform $h(t-\Delta t\coa)$, which is itself expressed in the source frame and
thus naturally yields the rest-frame frequency.

A crucial finding of this work is that the last term in
Equation~\eqref{eq:ho_deriv} can be expressed as
\begin{equation}
		\begin{split}
			&\fourier{\left(t-t\coa\right)^2 \pdv{h(t - \Delta t\coa)}{t}} \\
			=& \int_{-\infty}^{\infty} \left(t-t\coa\right)^2 \pdv{h(t - \Delta t\coa)}{t}\ee^{-i\omega t}\dd{t} \\
			=& -\int_{-\infty}^{\infty} \pdv{h(t - \Delta t\coa)}{t} \pdv[2]{\omega}(\ee^{-i\omega (t-t\coa)}) \ee^{-i\omega t\coa} \dd{t} \\
			=& - \ee^{-i\omega t\coa} \pdv[2]{\omega}\int_{-\infty}^{\infty} \pdv{h(t - \Delta t\coa)}{t}\ee^{-i\omega (t-t\coa)} \dd{t} \\
			=& - \ee^{-i\omega t\coa} \pdv[2]{\omega}(\ee^{i\omega t\coa}\fourier{\pdv{h(t - \Delta t\coa)}{t}}) \\
			=& - \ee^{-i\omega t\coa} \pdv[2]{\omega}(\ee^{i\omega \tau\coa}\fourier{\pdv{h(t)}{t}}).
		\end{split}
\end{equation}
In the last line, for the Fourier transformation of $\pdv*{h(t)}{t}$, 
if we integrate it by parts, we can derive
$\fourier{\pdv*{h(t)}{t}}= i\omega \tilde{h}$.
This finding allows us to write the observed waveform in the frequency domain as
\begin{equation}
		\tilde{h}\obs \simeq \tilde{h}\ee^{-i\omega \Delta t\coa} -ia\ee^{-i\omega t\coa} \pdv[2]{\omega}(\ee^{i\omega \tau\coa}\omega \tilde{h}).
\end{equation}
Moreover, in practice we can choose the time of coalescence of the source, $\tau\coa$, as our reference time $t\coa$,
and set both of their values to $0$. 
Subsequently, the waveform in the frequency domain can be further simplified. The deviation from
the standard vacuum template, induced by the line-of-sight acceleration, then takes the following concise form,
\begin{equation}
		\Delta \tilde{h} \simeq -i a \pdv[2]{\omega}(\omega \tilde{h}). \label{eq:deviation1st}
\end{equation}

We refer to this method of calculating the accelerated waveform as the
``frequency-domain spectral differentiation'' (FSD).  We emphasize that it does
not rely on the stationary-phase or post-Newtonian approximations. Therefore, this method
can be used to process the full IMR waveform. 

Since GW observation is more sensitive to phase changes rather than amplitude
changes, it is helpful to introduce an equivalent phase shift $\Delta\Psi$
that is caused by acceleration. Such a quantity also enable us to compare our results with the phase shifts
derived by earlier works \citep{vijaykumarWaltzingBinariesProbing2023a, lazarowGravitationalWaveformModel2024a}. 
Using the relationship
\begin{equation}
		\tilde{h}\obs = A\tilde{h}\ee^{i\Delta \Psi} = \tilde{h} + \Delta \tilde{h},
\end{equation}
where $A$ is the factor for amplitude change,
%and $\Delta\Psi$,
and assuming $A\simeq1$ and $\Delta\Psi \ll 1$, 
we find that
\begin{equation}
		\Delta\Psi \simeq \frac{\Im{\Delta \tilde{h}/\tilde{h}}}{1+ \Re{\Delta \tilde{h}/\tilde{h}}}, \label{eq:delta psi}
\end{equation}
where $\Im{\cdots}$ and $\Re{\cdots}$ denote, respectively, the imaginary and real parts of the quantity.

\section{Waveform Accuracy} \label{sec:result}
	
In this section, we compare the accelerated waveforms generated by different methods.
We only consider the case where the spins of the two
merging black holes (BHs) align with the orbital angular momentum.  Therefore,
we can neglect the precession effect and use the IMRPhenomD model
\citep{khanFrequencydomainGravitationalWaves2016} to generate the vacuum
non-accelerated waveform.  When calculating the accelerated waveform with our
FSD method, we have to apply Equation~\eqref{eq:deviation1st}. We analytically
derived the derivatives of each order of the IMRPhenomD waveform, thereby
avoiding undesirable behaviors that occur at the splicing points of the
IMRPhenomD waveform. Moreover, we consider two PSDs when calculating the
mismatch, which are \texttt{aLIGOZeroDetHighPower} for advanced LIGO (aLIGO)
and \texttt{EinsteinTelescopeP1600143} for the Einstein Telescope (ET). The
inclusion of the ET PSD allows us to access the feasibility of applying our
method to third-generation detectors
\citep{abbottExploringSensitivityNext2017}. 

When numerically generating the waveforms, we choose a sample rate of
$f_\mathrm{s}=\SI{4096}{\hertz}$. We also choose a low starting frequency,
$f_\mathrm{start}=\SI{5}{\hertz}$, to produce long waveforms. In the
calculation of mismatch, however, we choose a slightly higher cutoff frequency,
i.e., $f_\mathrm{low}=\SI{20}{\hertz}$ for aLIGO, and
$f_\mathrm{low}=\SI{10}{\hertz}$ for ET. This choice is made to avoid the
noise-dominated regions in the PSDs. 
	
\subsection{Phase Shift} \label{sec:phase shift}

We now compare the acceleration-induced phase shifts obtained from different
methods.  The TDS phased shift is derived following the steps detailed in
Section~\ref{sec:tds}.  The FSD one is derived according to
Equations~\eqref{eq:deviation1st} and \eqref{eq:delta psi}.  To compare with
previous works, we also compute the
phase shift following the conventional SPA and PN methods. For 
this purpose, we adopt Equation~(4) in
\cite{lazarowGravitationalWaveformModel2024a}, which includes corrections up to
the 3PN order as well as the first-order spin-orbit coupling term.

It is important to note that the initial phase of each waveform remains a free
parameter. To eliminate this freedom, we align the waveforms such that the
phase is defined to be zero at the instant the binary reaches the innermost
stable circular orbit (ISCO). The corresponding orbital frequency in the
rest frame of the source is given by
\begin{equation}
f_\mathrm{ISCO} = \frac{1}{6^{3/2}\uppi (m_1+m_2)},
\label{eq:fisco}
\end{equation}
where $m_1$ and $m_2$ are the masses of the two BHs.
We emphasize that this moment does not necessarily coincide with the transition
between the inspiral and intermediate regions in phenomenological waveform
models.

\begin{figure*}[]
\includegraphics[width=\textwidth]{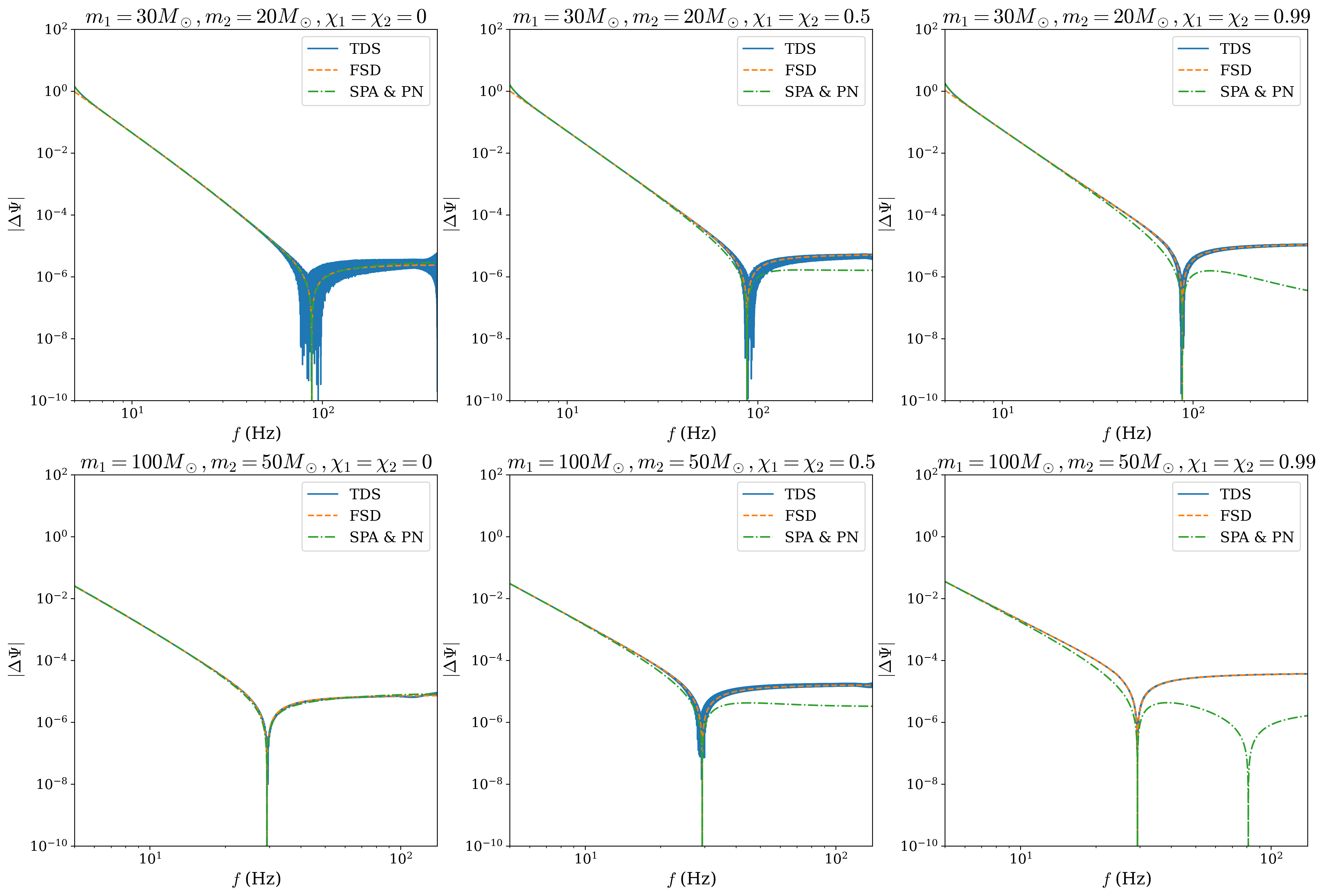}
\caption{Phase shifts as a function of frequency derived from different methods, including
the time domain stretching (TDS, blue solid line), frequency-domain spectral differentiation (FSD, orange dashed line),
and the SPA$+$PN method (green dot dashed line). Here we have assumed an effective acceleration of $a=\SI{e-5}{\per\s}$,
a cosmological redshift of $0.2$ for the BBH, and the standard $\upLambda$CDM cosmology. The other parameters
used in the calculation are labeled at the top of each panel.
}
\label{fig:delta psi}
\end{figure*}

Figure~\ref{fig:delta psi} compares the phase shifts obtained using different
methods. The TDS result, being exact, serves as the reference. The dip observed
in each TDS curve corresponds to the moment the BBH reaches the ISCO.  In
general, both the FSD and the SPA with PN correction (SPA+PN) methods reproduce
the TDS result reasonably well in the non-spinning case. However, when spin is
included, the phase shift from the SPA+PN method begins to deviate noticeably
from the TDS benchmark. This discrepancy becomes particularly prominent during
the merger and ringdown stages (to the right of the dip), highlighting the
breakdown of the stationary-phase and PN approximations in the strong-field
regime.  In contrast, the FSD method maintains high accuracy throughout the
inspiral, merger, and ringdown phases, even in the presence of spin. These
results demonstrate the robustness of the FSD approach in modeling the
late-stage evolution of accelerating BBHs.

\subsection{Mismatch} \label{sec:mismatch}

The difference between two waveforms can also be quantified by a ``mismatch'', which is defined as
\begin{equation}
		\mathcal{M} \equiv 1 - \max_{t_0,\phi_0}\frac{\braket{\tilde{h}_1}{\tilde{h}_2}}{\sqrt{\braket{\tilde{h}_1}{\tilde{h}_1}\braket{\tilde{h}_2}{\tilde{h}_2}}}, \label{eq:mismatch}
\end{equation}
where
\begin{equation}
		\braket{\tilde{h}_1}{\tilde{h}_2} = 2\int_{0}^{\infty}\frac{\tilde{h}_1(f)\tilde{h}_2^\ast(f) + \tilde{h}_1^\ast(f)\tilde{h}_2(f)}{S_\mathrm{n}(f)}\dd{f} \label{eq:innerprod}
\end{equation}
is the noise-weighted inner product of two different waveforms $\tilde{h}_{1}$ and $\tilde{h}_{2}$,
and $S_\mathrm{n}(f)$ is the one-sided power spectral density (PSD) of the noise. 
The maximum value for Equation~\eqref{eq:mismatch} is determined by optimizing the
starting times and the initial phases of the waveforms.

\begin{figure*}[t]
\includegraphics[width=\textwidth]{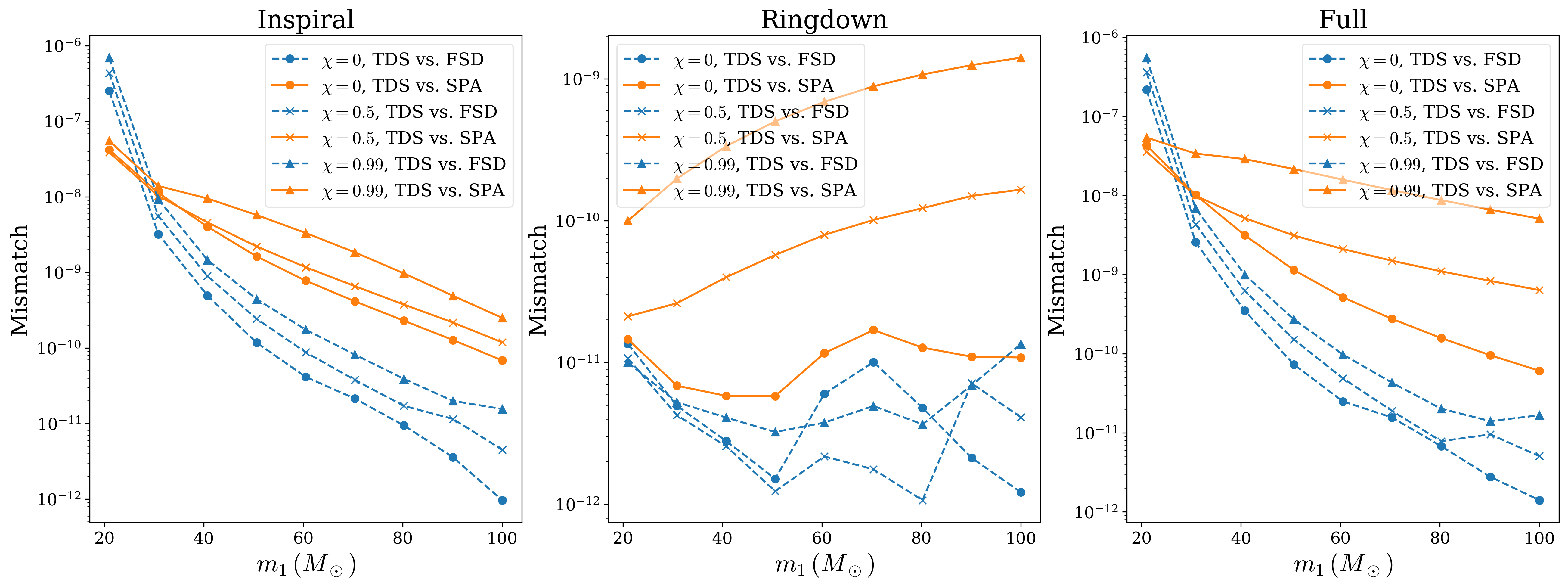}
\caption{Mismatch versus the mass of the primary BH. The orange solid curves 
show the mismatches between the TDS waveforms and SPA+PN ones, while the blue dashed curves refer to
the mismatches relative to the FSD waveforms.
Different symbols indicate different spin parameters, as detailed in the legend.
All configurations assume a mass ratio of $m_1:m_2 = 2:1$, spin parameters of $\chi_1 = \chi_2 = \chi$, and an effective acceleration of $a =
\SI{e-4}{\per\s}$, and the aLIGO sensitivity curve. 
	The left, middle,
and right panels display the mismatches for the inspiral,
merger--ringdown, and full IMR signal, respectively.
}
\label{fig:td vs dev vs pn ligo}
\end{figure*}

Figure~\ref{fig:td vs dev vs pn ligo} shows the mismatch between different
accelerated waveforms.  
Here, we have assumed the aLIGO sensitivity and an effective acceleration of
$a =\SI{e-4}{\per\s}$.
Using the
full waveforms (right panel), which include the inspiral, merger, and ringdown
phases, we observe that the mismatch decreases with increasing BH mass.
Notably, the FSD waveforms perform better than the SPA+PN ones for primary
masses $m_1 \gtrsim 30M_\odot$. This can be understood from the fact that
waveforms become shorter for higher masses, which better satisfies the
assumption $|a(t - t_\mathrm{coal})|^2 \ll f^{-1}$ used in the derivation of
the FSD waveform. Additionally, waveforms for spinning BHs, whether generated
by FSD or SPA+PN, generally exhibit larger mismatches relative to the TDS
benchmark, reflecting the added complexity of strong-gravity dynamics in the
presence of spin.

To assess the waveform accuracy particularly in the merger--ringdown phase, we apply
a cutoff at the ISCO frequency $f_\mathrm{ISCO}$ (Eq.~\eqref{eq:fisco}) and
separate the mismatch into contributions from the inspiral (left panel) and
merger--ringdown (middle panel) stages. The similarity between the inspiral and
full mismatches indicates that the SNR is mostly determined by the
inspiral stage, which dominates the signal duration. However, when focusing
solely on the merger--ringdown phase, we find that the FSD model consistently
yields a smaller mismatch than the SPA+PN model. This result further highlights
the limitations of the SPA and PN approximations in the strong-gravity regime
and confirms the effectiveness of our FSD approach in modeling the late-stage
evolution of accelerating BBHs.

The aforementioned limitation in inspiral-stage accuracy (for both FSD and SPA+PA waveforms) becomes more prominent
when considering the sensitivity of the ET. Owing to its lower noise floor at
low frequencies, ET is expected to detect inspiral signals of significantly
longer duration than aLIGO, a regime where the accuracy of our FSD method
currently deteriorates. This is illustrated in Figure~\ref{fig:td vs dev vs pn
1e-4}, which shows the mismatches for an effective acceleration of
$a=\SI{e-4}{\per\s}$.  While the FSD waveform maintains higher accuracy than
the SPA+PN waveform in the merger--ringdown stage, its inspiral-stage mismatch
increases more substantially, exceeding the mismatch of SPA+PN 
waveforms when $m_1\lesssim 70M_\odot$.

\begin{figure*}[]
\includegraphics[width=\textwidth]{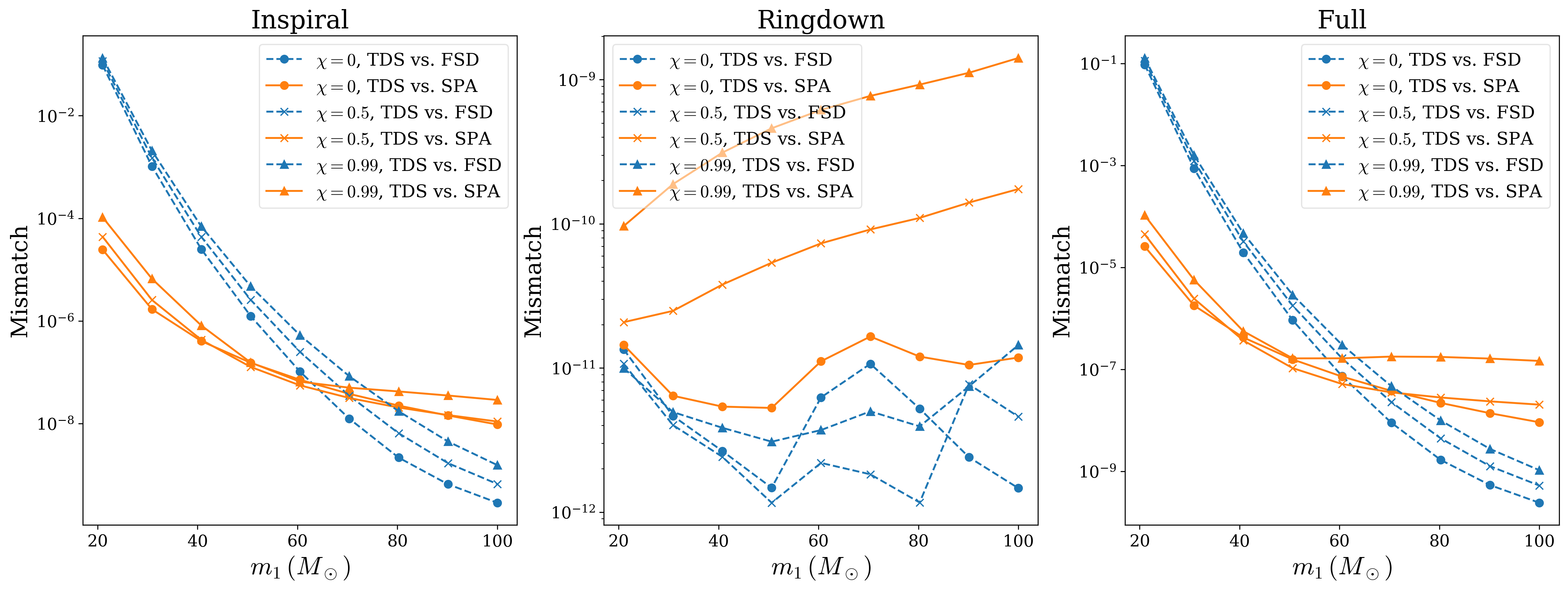}
\caption{The same as Figure~\ref{fig:td vs dev vs pn ligo} but using the ET sensitivity. The duration
of the waveforms are now longer due to the better sensitivity of ET.}
	\label{fig:td vs dev vs pn 1e-4}
\end{figure*}

The issue of signal duration is alleviated when the effective acceleration is
reduced to $a = \SI{e-5}{\per\s}$, as shown in Figure~\ref{fig:td vs dev vs pn
1e-5} (also for ET). In this case, the mass threshold, above which FSD performs 
systematically better than SPA+PN,
decreases slightly to
about $m_1 \simeq 40M_\odot$. Notably, the inspiral-stage mismatch of our FSD
method drops by three orders of magnitude as the acceleration decreases from
$10^{-4}$ to $10^{-5}~\text{s}^{-1}$. However, in this low-acceleration regime,
the accuracy advantage of FSD in the merger--ringdown stage becomes less
pronounced, particularly for waveforms incorporating spin, as seen in the
middle panels of Figures~\ref{fig:td vs dev vs pn 1e-4} and \ref{fig:td vs dev
vs pn 1e-5}. This is expected, as the physical imprint of acceleration
systematically weakens with decreasing $a$.

\begin{figure*}[]
		\includegraphics[width=\textwidth]{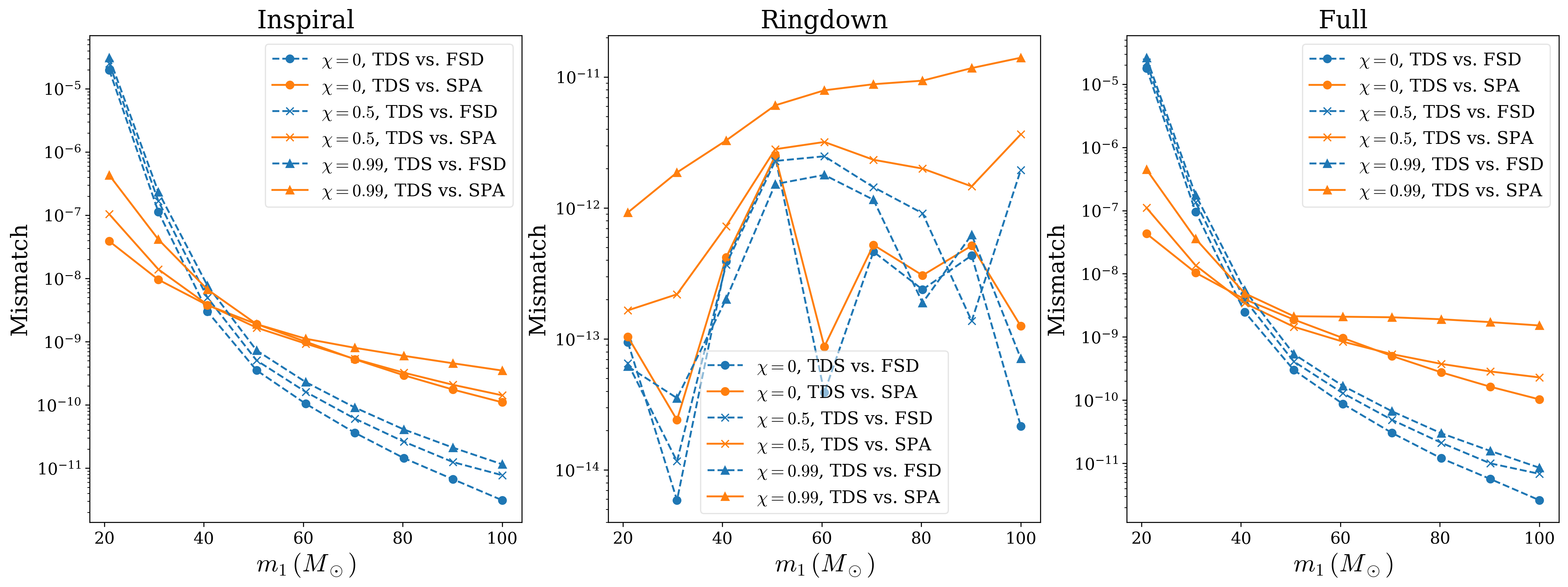}
		\caption{The same as Figure~\ref{fig:td vs dev vs pn 1e-4} but reducing the effective acceleration to $a=\SI{e-5}{\per\s}$.}
		\label{fig:td vs dev vs pn 1e-5}
\end{figure*}

\subsection{Higher Order Corrections to FSD waveforms}\label{sec:higher order}

The accuracy of the FSD method in the inspiral stage can be improved by
including higher-order terms in the expansion of Equation~\eqref{eq:deviation1st}.
In principle, as long as the condition $|a(t - t_\mathrm{coal})^2| < f^{-1}$ is
satisfied, higher-order expansions yield more accurate approximations.

In Appendix~\ref{sec:higher order derivation}, we derive an accelerated
waveform model based on the FSD method that can be expanded to arbitrary order
in $a(t - t_\mathrm{coal})^2$. Correspondingly, the deviation from the non-accelerated waveform
is given by
\begin{equation} 
	\Delta \tilde{h} \simeq \sum_{k=1}^{n}
\frac{a^k}{k!} (-i)^k \pdv[2k]{\omega} \left( \omega^k \tilde{h} \right).
\label{eq:deviation} 
\end{equation}

We use this expression to construct accelerated waveforms and evaluate their
mismatches against the reference TDS result. Figure~\ref{fig:higer order
comparison inspiral} shows the mismatch between TDS and FSD waveforms with
different expansion orders, along with the TDS-vs-SPA mismatch (red curve) for
reference. All configurations assume non-spinning binaries ($\chi_{1,2} = 0$)
and an effective acceleration of $a = \SI{e-5}{\per\s}$.

We find that while the first-order FSD expansion performs worse than the SPA+PN
model at the low-mass end, its accuracy improves significantly at second order
and surpasses SPA+PN when third-order terms are included. In the high-mass
regime, higher-order expansions bring little improvement, as the signal
duration is short enough that the leading-order deviation remains
adequate. In the merger--ringdown stage, results for different expansion orders
overlap, since this phase is too short to be significantly affected by
higher-order acceleration corrections.

\begin{figure*}[]
	\includegraphics[width=\textwidth]{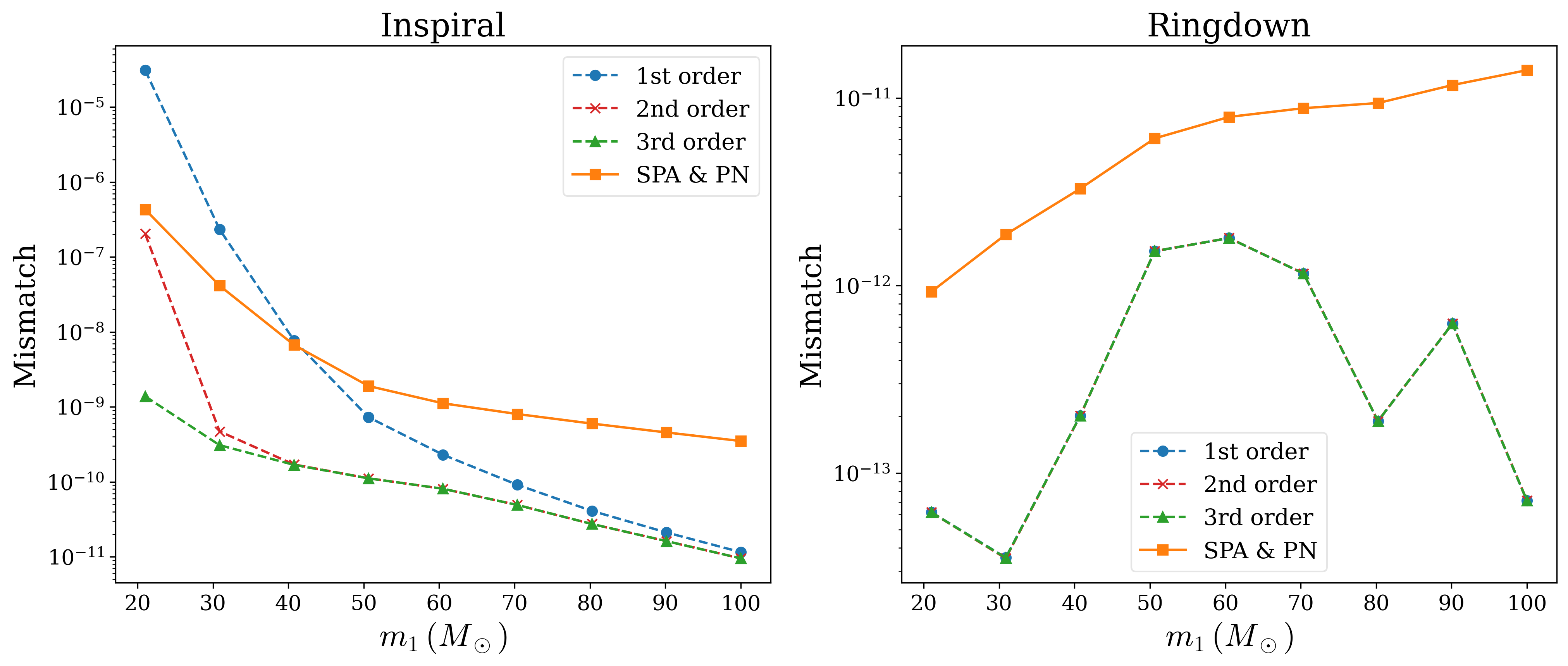}
	\caption{Similar to Figure \ref{fig:td vs dev vs pn 1e-5} but including different orders of corrections
	in the FSD waveform. Notice that in the merger--ringdown stage, the mismatches resulting from
	different orders of corrections overlap, because the duration of this stage is short.
	}
	\label{fig:higer order comparison inspiral}
\end{figure*}

\section{Constraints on Effective Acceleration} \label{sec:delta_a}

Now we estimate the precision of measuring the effective acceleration 
by future ground-based detectors, such as ET. We quantify the precision using the Fisher information matrix (FIM) method \citep{vallisneriUseAbuseFisher2008, lazarowGravitationalWaveformModel2024a}. The FIM is defined as
\begin{equation}
\Gamma_{ij} = \left\langle \pdv{\tilde{h}}{\theta_i} \middle| \pdv{\tilde{h}}{\theta_j} \right\rangle,
\end{equation}
and the variance of parameter $\theta_i$ is approximated by $\sigma_i^2 =
(\Gamma^{-1})_{ii}$. In this analysis, we treat the effective acceleration $a$
as the only free parameter, with all other binary black hole properties held
fixed. Although this setup is idealized, it enables a consistent comparison of
measurement precision across different waveform models.

\begin{figure}[]
\includegraphics[width=\columnwidth]{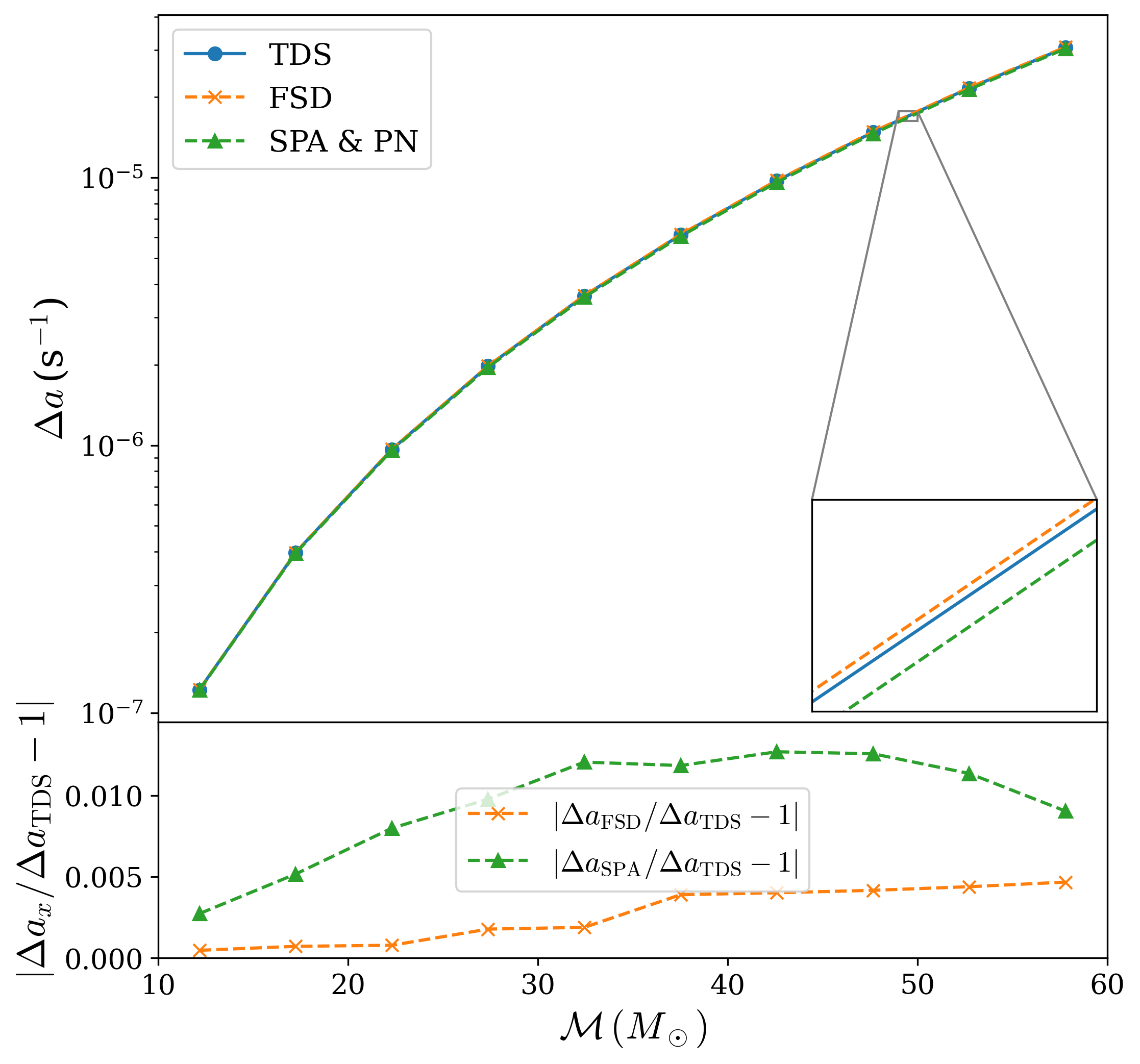}
	\caption{Precision in estimating the effective acceleration,
	evaluated using different waveform models. The
	measurement uncertainty is shown as a function of the
	chirp mass of BBH. All
	configurations assume a signal-to-noise ratio (SNR) of $1000$, a mass
	ratio of $m_1:m_2 = 2:1$, and non-spinning components ($\chi_{1,2}=0$).
	The top panel compares the absolute precision achieved by each waveform
	model, while the bottom panel shows the relative difference compared to
	the time-domain stretching (TDS) waveform results.
	} \label{fig:delta_a}
\end{figure}

Figure~\ref{fig:delta_a} shows the constraints on $a$ as a 
function of chirp mass $\mathcal{M} = (m_1
m_2)^{3/5}/(m_1+m_2)^{1/5}$. We adopt the third-order FSD-expanded waveform (as
established in Sec.~\ref{sec:higher order}) and fix the signal-to-noise ratio
at $\mathrm{SNR} = 1000$, a realistic value for ET
\citep{abbottExploringSensitivityNext2017}.

The upper panel compares the constraints derived from different waveform
models. Although all methods yield similar results, the FSD-based constraint
aligns more closely with the benchmark TDS result than the SPA+PN method does.
This is further quantified in the lower panel, which plots the relative
deviation from the TDS value, $|\Delta a_x / \Delta a_\mathrm{TDS} - 1|$,
confirming the better accuracy of the FSD approach.

Across the entire chirp mass range considered, the FSD method agrees with the
TDS result to within $0.5\%$—significantly better than the SPA+PN method. While
the quantitative improvement in precision is modest, the FSD method offers an
important conceptual advantage: it does not rely on the stationary phase or
post-Newtonian approximations, thus providing a more flexible and
waveform-agnostic framework suitable for future numerical-relativity-based
templates that may lack analytical expressions.

Before proceeding further, we would like to mention several cautions when applying the FSD method. 
Due to the nature
of the Fourier transformation, frequency-domain waveforms exhibit strong
oscillations at low frequencies. Sampling at rates significantly above
the Nyquist frequency of the time-domain signal is recommended to avoid
numerical instability. Furthermore, when computing high-order derivatives
numerically, particularly fourth order and beyond, instabilities may arise
because phenomenological waveforms are often spliced from multiple
segments that ensure continuity only up to the first derivative. To mitigate
this, we recommend using analytical derivatives where available or applying a
Gaussian filter to smooth numerical derivatives at segment boundaries. Finally,
since the first-order expansion already provides high accuracy in the
merger--ringdown stage, we advise restricting to first order in this stage while
employing higher orders only in the inspiral phase.

\section{Discussion and Conclusion} \label{sec:discussion}

In this work, we have proposed the frequency-domain spectral differentiation
(FSD) method for directly constructing gravitational waveforms in the frequency
domain for accelerating BBHs.  The modification in the frequency domain is
given in Equation~\eqref{eq:deviation1st} and generalized to arbitrary order in
Equation\eqref{eq:deviation}.  Our tests based on IMRPhenomD waveform confirms
that the FSD method, with a sufficiently high-order expansion, achieves higher
accuracy than the traditional SPA+PN approach.
Besides accuracy, the FSD method offers two other advantages.

First, regarding its regime of
validity, the FSD method only requires that the effective acceleration be
sufficiently small and the signal duration short enough to satisfy $|a(t -
t_\mathrm{coal})^2| \ll f^{-1}$. As demonstrated by
Equation~\eqref{eq:Newtonian_a} and the subsequent discussions, this condition
is generally met for stellar-mass compact binaries located near SMBHs,
particularly when the binaries enter the high-frequency band of ground-based
detectors.  This ensures that our approach is applicable to various types of
astrophysical scenarios discussed in Section~\ref{sec:intro}.

Second, the model-agnostic nature of the FSD method enables one to incorporate
acceleration effects into the most advanced and complex waveform templates. As
shown in previous works, orbital eccentricity and spin precession play a
critical role in breaking the parameter degeneracy between the environmental
Doppler shift and the binary's intrinsic chirp mass \cite{xuanDetectingAcceleratingEccentric2023,tiwariPipelineSearchSignatures2025,hendriksGravitationalWaveParameter2026a}.
However, deriving analytical SPA corrections for eccentric and precessing
systems can be mathematically intractable. Since the FSD method acts directly
on the final frequency-domain waveform, it can be applied to the
state-of-the-art templates (e.g., the IMRPhenom or effective-one-body families
\citep{prattenComputationallyEfficientModels2021,ramos-buadesNextGenerationAccurate2023},
accommodating subdominant modes, precession, as well as recent eccentric
waveform extensions
\citep{gamboaThirdPostNewtonianDynamics2025,ramos-buadesFastFrequencydomainPhenomenological2026}).

In terms of practical implementation, although analytical derivatives are
employed in this work with the IMRPhenomD model to showcase the precision, the FSD
method does not rely on analytical differentiation, and can equally work with
numerical differentiation. Crucially, even when employing numerical
derivatives, traditional TDS method, which relies on FFT, incurs a
computational cost that scales as $\mathcal{O}(n\log n)$, while the numerical
differentiation for the FSD method scales linearly as $\mathcal{O}(n)$ because
it performs purely point-wise differentiation in the frequency domain. This
reduction in computational time is vital for processing
high-sampling-rate signals in the high-frequency band or for long-duration
binary systems, where $n$ is exceptionally large. Furthermore, in
computationally intensive tasks such as Bayesian inference, Markov Chain Monte
Carlo (MCMC) or nested sampling algorithms typically require millions of
waveform likelihood evaluations. The speedup in single-waveform generation
provided by FSD reduces the overall analysis time, making it a
compelling tool for routine GW data analysis.

Beyond the improvement in computational efficiency, the FSD method
fundamentally resolves the physical inconsistency inherent in previous
accelerated waveform constructions. Because the SPA and PN formalism break down
during the non-adiabatic merger--ringdown stages, previous acceleration-induced phase
corrections are typically applicable only to the inspiral waveforms.
By contrast, the FSD method ensures that the entire IMR
evolution remains tractable within the same accelerated reference frame.

The ability of the FSD method to accurately extend into the merger--ringdown
stages has profound implications for fundamental physics. Precision tests of GR
rely heavily on the exact late-stage waveform. For instance, in recently
detected high-SNR events such as GW250114, potential signatures of non-linear
quadratic quasi-normal modes have been observed in the ringdown signal
\citep{yangContributionNonlinearQuasinormal2025,wangNonlinearVoiceGW2501142026}.
However, if the center-of-mass effective acceleration induced by the
environment is neglected during these highly non-linear stages, the resulting
kinematic frequency shifts can easily mimic such non-linear gravitational
effects or deviations from vacuum GR \cite{canevasantoroFirstConstraintsCompact2024,royCompactBinaryCoalescences2025,guptaPossibleCausesFalse2025}.
Traditional SPA+PN methods break down during the extremely
brief and non-adiabatic ringdown phase, while the FSD framework can operate
directly in the frequency domain, thereby effectively capture environmental
systematic errors and prevent them from contaminating strong-field GR tests.

It is important to emphasize that the FSD derivation assumes a constant
effective line-of-sight acceleration. This is a reasonable approximation for
the BBHs in the sensitive band of ground-based detectors, where the
acceleration varies slightly over the observable signal duration. However, for
those BBHs that may be observed by space-based detectors
\citep{sesanaProspectsMultibandGravitationalWave2016,kyutokuConciseEstimateExpected2016,bonettiPostNewtonianEvolutionMassive2019},
a periodic acceleration model
would be more appropriate since the duration of the signal could last months to
years.  Extending the FSD method to such scenarios will be useful for elucidating
the environments of the GW sources for space detectors.

\begin{acknowledgments}
	This work is supported by the National Key Research and Development Program of China (Grant No. 2021YFC2203002).
X.C. is also supported by the National Natural Science Foundation of China (Grant No. 12473037).
\end{acknowledgments}
	
	\bibliography{main.bib}
	\bibliographystyle{apsrev4-2}

	\appendix

\section{Higher-order corrections to the FSD method} \label{sec:higher order derivation}

The accelerated waveform in the frequency domain can be more accurately derived by 
expanding the time-domain signal $h\obs(t)$ to the $n$-th order of $a(t-t\coa)^2$ and then performing the
Fourier transformation. Consequently,  the observed signal in the time domain becomes
\begin{equation}
		h\obs(t) \simeq h(t - \Delta t\coa) + \sum_{k=1}^{n} \frac{a^k}{k!} \left(t-t\coa\right)^{2k} \pdv[k]{h(t-\Delta t\coa)}{t}.
	\end{equation}
Performing a Fourier transformation, the frequency-domain waveform becomes
\begin{equation}
		\tilde{h}\obs(f) \simeq \tilde{h}(f)\ee^{-i\omega \Delta t\coa} + \sum_{k=1}^{n} \frac{a^k}{k!} \fourier{\left(t-t\coa\right)^{2k} \pdv[k]{h(t-\Delta t\coa)}{t}}.
\end{equation}
We can calculate the $k$th-order term in the previous Fourier transformation by
\begin{equation}
		\fourier{\left(t-t\coa\right)^{2k} \pdv[k]{h(t-\Delta t\coa)}{t}} = (-i)^k \ee^{-i\omega t\coa} \pdv[2k]{\omega}(\ee^{i\omega \tau\coa} \omega^k \tilde{h}).
\end{equation}
Finally, the full correction to the frequency-domain waveform due to acceleration is
\begin{equation}
		\Delta \tilde{h} \simeq \sum_{k=1}^{n} \frac{a^k}{k!} (-i)^k \ee^{-i\omega t\coa} \pdv[2k]{\omega}(\ee^{i\omega \tau\coa} \omega^k \tilde{h}).
\end{equation}
By setting $t\coa=\tau\coa=0$, the correction factors reduce to
\begin{equation}
	\Delta \tilde{h} \simeq \sum_{k=1}^{n} \frac{a^k}{k!} (-i)^k \pdv[2k]{\omega}(\omega^k \tilde{h}).
\end{equation}

\end{document}